\input harvmac
\noblackbox
 
\font\ticp=cmcsc10
 
\def\Title#1#2{\rightline{#1}\ifx\answ\bigans\nopagenumbers\pageno0\vskip1in
\else\pageno1\vskip.8in\fi \centerline{\titlefont #2}\vskip .5in}

\font\ticp=cmcsc10
\font\ttsmall=cmtt10 at 8pt

\input epsf
\ifx\epsfbox\UnDeFiNeD\message{(NO epsf.tex, FIGURES WILL BE
IGNORED)}
\def\figin#1{\vskip2in}
\else\message{(FIGURES WILL BE INCLUDED)}\def\figin#1{#1}\fi
\def\ifig#1#2#3{\xdef#1{fig.~\the\figno}
\goodbreak\topinsert\figin{\centerline{#3}}%
\smallskip\centerline{\vbox{\baselineskip12pt
\advance\hsize by -1truein\noindent{\bf Fig.~\the\figno:} #2}}
\bigskip\endinsert\global\advance\figno by1}

%
\def\p{\partial}
\def\[{\left [}
\def\]{\right ]}
\def\({\left (}
\def\){\right )}

\def\BH{black hole}
\def\GL{Gregory-Laflamme}
\def\GM{Gubser-Mitra}
\def\L{\Lambda}
\def\l{\ell}
\def\lt{\tilde L_q}
\def\lb{\bar L_r}
\def\sq{{\bf S}^q}
\def\sqr{{\bf S}^q \times {\bf S}^r}
\def\sp{{\bf {S}^p}}

\def\sr{{\bf S}^r}
\def\S{{\bf S}}
%
\lref\grla{
R.~Gregory and R.~Laflamme,
``Black strings and p-branes are unstable,''
Phys.\ Rev.\ Lett.\  {\bf 70}, 2837 (1993)
[hep-th/9301052]. 
R.~Gregory and R.~Laflamme,
``The Instability of charged black strings and p-branes,''
Nucl.\ Phys.\ B {\bf 428}, 399 (1994)
[hep-th/9404071].} 
\lref\gumi{
S.~S.~Gubser and I.~Mitra,
``Instability of charged black holes in anti-de Sitter space,''
hep-th/0009126.
S.~S.~Gubser and I.~Mitra,
``The evolution of unstable black holes in anti-de Sitter space,''
JHEP {\bf 0108}, 018 (2001)
[hep-th/0011127].}
\lref\reall{
H.~S.~Reall,
``Classical and thermodynamic stability of black branes,''
Phys.\ Rev.\ D {\bf 64}, 044005 (2001)
[hep-th/0104071].}
\lref\homa{
G.~T.~Horowitz and K.~Maeda,
``Fate of the black string instability,''
hep-th/0105111.}
\lref\prestidge{
T.~Prestidge,
``Dynamic and thermodynamic stability and negative modes in  Schwarzschild-anti-de Sitter,''
Phys.\ Rev.\ D {\bf 61}, 084002 (2000)
[hep-th/9907163].}
\lref\hika{
T.~Hirayama and G.~Kang,
``Stable black strings in anti-de Sitter space,''
Phys.\ Rev.\ D {\bf 64}, 064010 (2001)
[hep-th/0104213].}
\lref\grro{
J.~P.~Gregory and S.~F.~Ross,
``Stability and the negative mode for Schwarzschild in a finite cavity,''
hep-th/0106220.}
\lref\bousso{
R.~Bousso,
``Positive vacuum energy and the N-bound,''
JHEP {\bf 0011}, 038 (2000)
[arXiv:hep-th/0010252].
R.~Bousso,
``Bekenstein bounds in de Sitter and flat space,''
JHEP {\bf 0104}, 035 (2001)
[arXiv:hep-th/0012052].
R.~Bousso,
``Holography in general space-times,''
JHEP {\bf 9906}, 028 (1999)
[hep-th/9906022].}
\lref\holography{
R.~Bousso,
``The holographic principle for general backgrounds,''
Class.\ Quant.\ Grav.\  {\bf 17}, 997 (2000)
[hep-th/9911002].
R.~Bousso,
``Holography in general space-times,''
JHEP {\bf 9906}, 028 (1999)
[hep-th/9906022].
R.~Bousso,
``A Covariant Entropy Conjecture,''
JHEP {\bf 9907}, 004 (1999)
[hep-th/9905177].}
\lref\dSstab{
J.~C.~Niemeyer and R.~Bousso,
``The nonlinear evolution of de Sitter space instabilities,''
Phys.\ Rev.\ D {\bf 62}, 023503 (2000)
[gr-qc/0004004].}
\lref\hust{
G.~J.~Stephens and B.~L.~Hu,
``Notes on black hole phase transitions,''
gr-qc/0102052.}
\lref\membrane{
K.~S.~Thorne, R.~H.~Price and D.~A.~Macdonald,
``Black Holes: The Membrane Paradigm,''
{\it  New Haven, USA: Yale Univ. Pr. (1986) 367p}.}
\lref\qnm{
G.~T.~Horowitz and V.~E.~Hubeny,
``Quasinormal modes of AdS black holes and the approach to thermal  equilibrium,''
Phys.\ Rev.\ D {\bf 62}, 024027 (2000)
[hep-th/9909056].}
\lref\chandra{
S.~Chandrasekhar, {\it The Mathematical Theory of Black Holes}, 
Oxford University Press, Oxford (1983).}
\lref\gpy{D.~J.~Gross, M.~J.~Perry and L.~G.~Yaffe,
``Instability of flat space at finite temperature,''
Phys.\ Rev.\ D {\bf 25}, 330 (1982).}
\lref\dewolfe{
O.~deWolfe, D.~Z.~Freedman, S.~S.~Gubser, G.~T.~Horowitz and I.~Mitra,
``Stability of $AdS_p \times M_q$ compactifications without supersymmetry,''
hep-th/0105047.}
\lref\krn{
H.~.J.~Kim, L.~J.~Romans and P.~van Nieuwenhuisen,
``Mass spectrum of chiral N=2 D = 10 supergravity on ${\bf S^5}$,''
Phys.\ Rev.\ D {\bf 32}, 389 (1985).}
\lref\witheads{
E.~Witten,
``Anti-de Sitter space, thermal phase transition, and confinement in
gauge theories,''
Adv.\ Theor.\ Math.\ Phys.\  {\bf 2}, 505 (1998)
[arXiv:hep-th/9803131].}
\lref\gubserfo{
S.~S.~Gubser,
``On non-uniform black branes,''
arXiv:hep-th/0110193.}
\lref\gli{
R.~Gregory and R.~Laflamme,
``Hypercylindrical black holes,''
Phys.\ Rev.\ D {\bf 37}, 305 (1988).}
\lref\emre{
R.~Emparan and H.~S.~Reall,
``A rotating black ring in five dimensions,''
arXiv:hep-th/0110260.}
\lref\wald{
R.~Wald, private communication.}
\lref\juan{J. Maldacena, 
``The large N Limit of superconformal field theories and supergravity,''
Adv. Theor. Math. Phys. 2 (1998) 231, hep-th/9711200.}
\lref\witten{ E. Witten, ``Anti de Sitter space and holography,''
Adv. Theor. Math. Phys. 2 (1998) 253, hep-th/9802150.}
\lref\gkp{S. Gubser, I. Klebanov, and A. Polyakov, 
``Gauge theory correlators from non-critical string theory,''
Phys. Lett. B428 (1998) 105,
hep-th/9802109.}
\lref\magoo{For a comprehensive review, see O. Aharony, S.S. Gubser,
J. Maldacena, H. Ooguri, and Y. Oz,
``Large N field theories, string theory and gravity",
hep-th/9905111.}
\lref\hapa{S. Hawking and D. Page, 
``Thermodynamics of black holes in Anti-de Sitter space,'' 
Commun. Math. Phys. 87 (1983) 577.}
\lref\homach{
G.~T.~Horowitz and K.~Maeda,
``Inhomogeneous near-extremal black branes,''
arXiv:hep-th/0201241.}
\lref\cawa{
S.~M.~Carroll, J.~Geddes, M.~B.~Hoffman and R.~M.~Wald,
``Classical stabilization of homogeneous extra dimensions,''
arXiv:hep-th/0110149.}
\lref\kangb{
G.~W.~Kang,
``On the stability of black strings / branes,''
arXiv:hep-th/0202147.}
\lref\kanga{
T.~Hirayama and G.~Kang,
``Stable black strings in anti-de Sitter space,''
Phys.\ Rev.\ D {\bf 64}, 064010 (2001)
[arXiv:hep-th/0104213].
}

%
%
\baselineskip 16pt
\Title{\vbox{\baselineskip12pt
\hbox{hep-th/0202189}\hbox{SU-ITP 02/08}\hbox{PUPT-2026}
}}
{\vbox{
{\centerline{Unstable Horizons}}
}}
\centerline{\ticp Veronika E. Hubeny$^1$ and Mukund Rangamani$^2$
\footnote{}{\ttsmall
 veronika@itp.stanford.edu, mukund@feynman.princeton.edu}}
\bigskip
\centerline{\it $^1$Physics Department, Stanford University,
Stanford, CA 94305, USA}
\bigskip
\centerline{\it $^2$ Department of Physics, Princeton University, Princeton, 
NJ 08544, USA}
\bigskip
\centerline{\bf Abstract}
\bigskip
We investigate some of the issues relating to the dynamical instability of 
general static spacetimes with horizons.
Our paper will be partially pedagogical and partially exploratory in nature.
After discussing the current understanding,
we generalize the proposal of Gubser and Mitra, which identifies dynamical 
instability of black branes with local thermodynamic instability, 
to include all product spacetimes with the horizon uniformly smeared over
an internal space.
We support our conjecture by explicitly exhibiting a threshold unstable mode
for Schwarzschild-$AdS_5 \times \S^5$ black hole.
Using the AdS/CFT correspondence,
this simultaneously yields a prediction for a phase transition 
in the dual gauge theory.
We also discuss implications for spacetimes with cosmological horizons.
\Date{February, 2002}

%
\newsec{Introduction}

Understanding the stability of a given spacetime is an important 
issue from many standpoints. 
In general relativity, stability suggests physical relevance, 
since a stable static spacetime may be 
approached dynamically from a reasonably generic set of initial conditions.
From a string theory perspective, it is 
interesting to know which spacetimes are good backgrounds for studying 
string propagation and which ones have tachyons in the stringy spectrum.
This issue gets further bolstered in light of the fact that
 in certain backgrounds, string propagation is holographically encoded 
by the dynamics of gauge theories ``living at the boundary'' of spacetime.
In such cases, the instability on the gravitational side is an indicator 
of interesting gauge theory dynamics, such as phase transitions, {\it etc.}.
Another interesting application relates to closed string 
tachyon condensation, since 
an instability in the gravitational background usually signals
a physical tachyonic mode in the string spectrum.

While the prospects of answering the question of stability of general spacetimes,
in a universal fashion, appear rather dim, 
it seems much more likely that there could be some underlying principle, 
which can be used 
for spacetimes which have horizons. 
Indeed, encouraged by the successes of such universal laws as those of \BH\ 
mechanics, 
one may be hopeful
that black hole spacetimes could be analyzed for (in)stability 
in a general setting. 
Therefore, the question we wish to pose is:
{\it What are the necessary and sufficient criteria for the onset of a dynamical 
instability of a horizon?}
Apart from deepening our understanding of the properties of horizons,
and thereby hopefully shedding more light on quantum gravity,
this has further uses in specific contexts.
For example, in the context of the AdS/CFT correspondence, one
obtains a better understanding of the behaviour of the gauge theory in a typical 
excited state and its phase transitions.
More generally, a universal criterion would not only obviate the need to 
carry out laborious linear analysis in order to determine stability, 
but it may be useful even in the cases where such analysis is impossible
to carry out because the metric is not explicitly known, 
but only the appropriate 
coarse-grained features are understood.

The question of black brane instability
was recently addressed by Gubser and Mitra 
\gumi, who presented a conjecture for configurations with non-compact
translational symmetry along the horizon. 
Namely, they proposed (and tested 
numerically) that such a black brane
is dynamically unstable to linearized perturbations when it is (locally)
thermodynamically unstable.
For spacetimes which are holographically dual to some field theory, this
is manifested as an instability of the ``thermal'' ensemble 
in the field theory.
More recently, the proposal received further evidence from Reall \reall,
who demonstrated the equivalence of the 
classical Lorentzian {\it threshold unstable mode}---%
which signals the onset of the  \GL\ instability---%
and the Euclidean negative mode
associated with thermodynamic instability.

While this seems to be a compelling story for the class of configurations 
considered by Gubser and Mitra, it is desirable to have a criterion 
which could address more general 
spacetimes in the absence of 
translational symmetries or non-compact $U(1)$ symmetries. Not only are 
the cases with compact $U(1)$s of interest, but also we would like to 
consider situations with more complicated symmetries of the extra directions,
such as
$SO(6)$, a situation 
pertinent for Schwarzschild Anti-de Sitter black hole which 
is smeared over a five sphere.

We will see that with a slight modification of the Gubser-Mitra conjecture
we can formulate a criterion which will allow us to talk about more 
general settings.  In particular, we will consider all product spacetimes,
wherein the black hole is smeared uniformly over some ``internal'' space.
However, we do not yet have a complete picture and 
our approach will be somewhat experimental in flavour. 
We will survey the 
known cases of Gregory-Laflamme like instabilities and attempt to glean from 
them lessons which are indicative of a pattern. 
This will lead to our proposal, which we will support 
using the particular case
of Schwarzschild-$AdS_5 \times \S^5$ black hole. 
We chose this particular case because 
of its special interest in the context of the AdS/CFT duality.
Our main result will lie 
in establishing the threshold instability for small Schwarzschild AdS black 
holes. 
This will simultaneously yield a prediction for the onset of a phase transition
in the dual gauge theory.

As further examination of our proposal, we will also briefly consider 
cosmological horizons.  Apart from the need to understand these better
within string theory, the intrigue in 
cosmological horizons stems from the fact that
they are observer-dependent.  Since even observer-dependent horizons 
have many properties in common with the genuine spacetime event horizons,
we wish to ask whether  observer-dependent horizons can also be subject
to \GL\ instabilities.  Of course, this requires the presence of some
transverse space.  We will consider spacetimes of the 
form $dS_p \times \S^q$ and $dS_p \times \S^q \times \S^r$.
While our conjecture suggests that the former is stable, the
surprising result will be that the latter is, in fact, marginally unstable
in a special case.
While these are neither phenomenologically relevant nor presently 
realized in string theory, they nevertheless may provide important lessons
for future considerations. 

The organization of the paper is as follows: 
In order to motivate our proposal,
we will start in Section 2 by reviewing the
known examples of Gregory-Laflamme instabilities.
We will then formulate a slightly modified version of the Gubser-Mitra conjecture
in Section 3, 
and test it in Section 4
by studying the example of AdS Schwarzschild black holes. 
Having established threshold instability for such systems, 
we will briefly comment on 
its implications for the dual 
gauge theoretic description.
We will then consider 
our conjecture in the context of cosmological horizons in Section 5, and 
 conclude in Section 6 with a summary and some speculations about more general 
spacetimes.


\newsec{Examples of Unstable Horizons}

Ideally, given a general spacetime with a
horizon, one would like to find and understand
a universal criterion for the horizon to be unstable to 
perturbations\foot{Of course, this assumes an affirmative
answer to the question ``is there a universal criterion?''
As mentioned earlier, the main reason for the authors' optimism stems 
from the lesson that horizons exhibit unexpectedly universal behaviour.}.
In the absence of some immediate general principle which would guide us,
we may consider the hints that already appear in the literature.
We have several ``data points'', {\it i.e.}, statements about specific \BH s or
classes of \BH s, which may provide clues to the generic case with arbitrary
event horizon.  
These usually involve a one-parameter family of solutions to Einstein's 
equation and specify the  value of the parameter which separates
the classically stable solutions from the unstable ones.
Rather than adhering to the chronological development of ideas, we 
will discuss these in order of increasing conceptual difficulty.

\subsec{Noncompact, translationally invariant cases}

Perhaps the simplest class of horizons to consider are those
exhibiting noncompact translational invariance, namely black branes.
Recently, a very concrete proposal has been suggested by Gubser and Mitra 
\gumi, for this subclass of event horizons.
The \GM\ conjecture states that for black branes with 
 noncompact translational symmetry, 
the \GL\ instability, {\it i.e.}, a tachyonic mode in small perturbations, is 
equivalent to a local\foot{
Gubser and Mitra use the terminology ``local'', to 
 distinguish it  from the global thermodynamic instability, 
given by the existence of a more entropically favorable configuration.}
 thermodynamic instability,
as manifested by the Hessian matrix
having positive eigenvalues.\foot{
Let us elaborate a little on this scenario. As \gumi, we will
work in the microcanonical ensemble, with fixed energy
or energy density. 
Then it is most natural 
thermodynamically to consider the second variation of the entropy with 
respect to energy or mass as the analog of specific heat, a concept 
familiar in the canonical ensemble. In general, one could also consider
chemical potentials and fix other charges; then the information about the 
stability of the ensemble is encoded in the corresponding Hessian matrix.
Specifically, for conserved quantities (including the energy) $\{ Q_i \}$,
the Hessian matrix is given by $H_{ij} \equiv {\p^2 S \over \p Q_i \p Q_j}$.
}
To test their conjecture, \gumi\  study particular
 plane-symmetric $AdS_4$-Reissner-Nordstr\" om black branes.
In the system considered,
the instability is manifested in the linearized approximation,
not by a growing mode of 
metric perturbations, but rather by a growing mode of the vector potential
perturbations, which are easier to handle computationally.
This gives evidence for the conjecture to a high numerical accuracy.

Further evidence for the conjecture was provided by Reall \reall,
whose argument relies on the fact that we can think of a translationally 
invariant black brane solution as a black hole which is smeared 
homogeneously along the extra directions. 
Furthermore, some 
black hole geometries, when thought of in Euclidean signature, have a 
negative eigenmode for the Euclidean Lichnerowicz operator. This mode typically
has no relevance for the dynamics of the black hole spacetime, because the 
Lorentzian equations are different from the Euclidean 
Lichnerowicz  operator eigenvalue equations. However, if we 
consider only static Lorentzian perturbations 
about the black string background, 
then we can reduce the Lorentzian problem to the Euclidean eigenvalue 
equation.  This is what Reall proposes to call the {\it threshold unstable 
mode}, since it is the mode which, in the Lorentzian geometry, signals the 
onset of the instability.

To give an explicit example,
consider the case of a black string in five-dimensions. 
This is just the 
four-dimensional Schwarzschild black hole
whose horizon is smeared along a non-compact translationally invariant
fifth direction. 
The 
corresponding Euclidean
continuation of the four-dimensional black hole 
has a mode with negative eigenvalue. 
This is precisely 
the mode discovered in the context of determining the stability of 
hot flat space, by Gross, Perry, and Yaffe \gpy. 
The eigenvalue equation for the Euclidean metric fluctuations is
\eqn\eucleqn{
\Delta_E h_{\mu \nu} = \lambda_E \, h_{\mu \nu}
}
with $\lambda_E < 0$. 
For time independent fluctuations in Lorentzian 
signature, the equations we obtain are just the same as the Euclidean 
ones, since $\Delta_L (static) = \Delta_E$.
 However, one does not associate any meaning to the negative eigenmode in 
the Lorentzian context, because the fluctuation equation corresponds to
the zero eigenvalue case: $\Delta_L h_{\mu \nu} = 0$.
So the Euclidean negative mode has no role to 
play in the dynamical evolution of the black hole spacetime. 

However, if we now have some extra directions transverse to the
black hole, as in the black string case, 
then we could consider fluctuations 
which carry momentum along the transverse directions.  Namely,
$h_{\mu \nu}(x,y) = h_{\mu \nu}(x) e^{ik y}$, where $x$ denotes the coordinate
in the black hole directions and $y$ denotes the transverse ${\bf R}$ coordinate.
Then the fluctuations are governed by the equation
\eqn\bstrliceqn{
\Delta_L h_{\mu \nu} (x) + k^2 \, h_{\mu \nu}(x) = 0.
}
This is now identical to the Euclidean eigenvalue equation \eucleqn, with 
the role of the eigenvalue being played by the momentum carried by the 
fluctuations in the transverse directions, $\lambda_E \equiv -k^2 $. 
Clearly, there exists a critical momentum, which 
coincides precisely with the eigenvalue one calculates in the 
Euclidean spacetime. This gives the  
threshold unstable mode.

The relation between the Euclidean negative mode of the \BH\ and the 
Lorentzian static mode of the black string was actually noticed much 
earlier by Gregory and Laflamme \gli.  Reall
\reall\ elaborates on this, and, working in the canonical ensemble,
shows that a black hole 
with negative specific heat cannot be stable and therefore must have a
Euclidean negative mode. 
Since the local thermodynamic instability 
is manifested by negativity of specific heat in the canonical ensemble, 
and equivalently by the positivity of the Hessian matrix
in the microcanonical ensemble, this explains the 
connection between classical (Lorentzian) stability and thermodynamic stability
of black branes conjectured by Gubser and Mitra \gumi.
Later, Gregory and Ross \grro, who considered neutral black branes in finite
cavity, provided a more complete argument
 that the Euclidean negative mode (and therefore the 
classical instability) occurs precisely when the black brane is 
thermodynamically unstable.
However, no complete proof of the universal equivalence between
the Euclidean negative mode and local thermodynamic instability
has been established so far.

While the \GM\ proposal yields a rather nice picture relating the 
thermodynamic and the classical dynamical instabilities, conceptually
it leaves something to be desired.  The former, manifested {\it e.g.},  
by the specific heat being negative, pertains to the dimensionally-reduced
black hole, whereas the \GL\ instability manifests itself by breaking
translational invariance in the extra dimensions, which one would 
associate rather with the properties of the transverse space.
With \gumi 's restriction to noncompact translationally invariant cases, 
the only ``property'' of the transverse space that one is free to vary 
is the dimensionality.  

Given the success of the conjecture for the noncompact 
translationally invariant branes, one may be tempted to try to generalize the
conjecture further by postulating a more universal applicability.
However, such naive approach fails at once, because of several known 
counter examples. 

\subsec{Compactified black string}

The earliest and best-known example of the \GL\ instability is the black 
string with a compact horizon, {\it i.e.},
the Schw$_4 \times {\bf S^1}$ geometry.
Specifically, in static coordinates, the metric is
\eqn\bs{
ds^2 = - \( 1 - {r_+ \over r} \) \, dt^2 +
       \( 1 - {r_+ \over r} \)^{-1} dr^2 + r^2 \, d \Omega_2^2
      + dz^2 ,}
where $z$ is identified with $z+L$, 
{\it i.e.} the length of the compact direction is $L$,
and the \BH\ horizon radius is $r_+$. 
For the noncompact $L \to \infty$ case,
\grla\ found that in  perturbing this geometry, all modes with sufficiently 
long wavelength (of the order of the 4-d \BH\ radius,
 specifically
$\lambda \ge 7.1 \, r_+$) are dynamically unstable.   
In the compactified case, the analysis is identical, except that now the 
effective masses of the higher modes will be quantized in units of
${2 \pi n \over L}$.
Equivalent picture, but one perhaps more useful conceptually, is to  
fix the length of the compact direction $L$ and vary $r_+$.
Then the \GL\ analysis shows that only
sufficiently small black holes will become dynamically
unstable to clumping in the compact direction.

At the first glance, this may seem inconsistent with the \GM\ conjecture,
since the 4-dimensional Schwarzschild \BH\ has 
positive Hessian matrix ${\p^2 S \over \p E^2} |_{r_+} = 8 \pi >0$, 
or equivalently, negative specific heat for all values of its radius, 
$C_v \equiv {\p E \over \p T} |_{r_+} = - 2 \pi L r_+^2 < 0$,
so that one would naively expect that all
\BH s should exhibit a corresponding dynamical instability.
In fact, there is no contradiction, simply because the conjecture does
not apply in this case:  while translational invariance is satisfied, the
extra direction is compact.  In the noncompact limit, corresponding to the
length of the $S^1$ getting large, $L \to \infty$,
the \GM\ proposal is satisfied, since all 
\BH s, having finite radius, are ``small'' ({\it i.e.},\ $r_+ \ll L$)
and therefore unstable.

This means that while there is a Euclidean negative mode for all values of 
the black hole radii $r_+$ 
(for a fixed $L$), 
only for sufficiently small $r_+$  
is this negative mode associated with a 
dynamical instability in Lorentzian time.  
It is a very interesting question, how  this Euclidean negative mode is
manifested in the absence of any dynamical instability.

Gregory and Laflamme 
also noted that the entropy of a fully 5-dimensional black hole\foot{  
{\it i.e.},\ where a spacial slice of the horizon has topology of $\S^3$,
rather than $\S^2 \times {\bf R}$, as for the black string.}
is greater
than that of the black string with same mass, when 
$L$ is large enough. 
This suggested that an instability of the kind found by their linearized
perturbation analysis may initiate the fragmentation of the horizon, so
that the black string pinches off and 
becomes a 5-dimensional black hole. 
Such a process would be entropically favorable
in roughly the same regime as where the 5-d \BH\ could ``fit'' in the
compact direction, namely for small \BH s.

This picture of black string fragmentation was widely accepted until 
recently, when Horowitz and Maeda \homa\
reexamined the fate of the instability,
 with very surprising and intriguing outcome: 
Instead of fragmenting into disconnected pieces, 
the horizon can only deform.  It cannot pinch off, because any $\S^2$
on the horizon cannot shrink to zero size\foot{
This was proved by \homa\ only for finite proper time using the
Raychaudhuri equation, but 
strong arguments were presented even for the infinite time case.
Their argument in fact holds for general horizons, which we will make 
use of later.}.
Given this, the conjecture for the black string is that it settles down to a
new static black string solution which is not translationally invariant
along the horizon. 
Nevertheless, (the converse of) the entropy argument 
may still be used as some indicator
of whether an instability is possible: when the entropy of the 5-d \BH\
is smaller than that of the corresponding black string configuration, 
we would not expect any deformation of the black string to be 
entropically favorable, so that the black string should be 
stable.

For completeness, we present
 the specific calculation of the transition point, {\it i.e.},\ the 
relation between $r_+$ and $L$, for which the black string and the 
 5-d \BH\ of the same mass also have the same entropy.
Let us approximate\foot{
This will not be the exact solution because of the finite size of the
compact direction.  Physically, one may think of this in the decompactified
picture as the horizon being deformed by an
infinite periodic array of \BH s.
However, the  area of the horizon should still be the same, by the following
(handwaving) argument:
The area is given by the entropy, which counts the number of the 
internal states.  Starting with the \BH s infinitely far apart, 
and therefore described by the exact solution, 
we wouldn't expect this number to change as we bring the \BH s
near each other.}
 the 5-d \BH\ by the ordinary Schw$_5$ metric,
with  5-d radial coordinate $R$ and the horizon radius $R_+$:
\eqn\bh{
ds^2 = - \( 1 - {R_+^2 \over R^2} \) \, dt^2 +
       \( 1 - {R_+^2 \over R^2} \)^{-1} dR^2 + R^2 \, d \Omega_3^2}
The masses of the black string and this 5-d \BH\ are, respectively,
$M_{bs} = {1 \over 2} r_+ L$ and $M_{bh} = {3 \pi \over 8} R_+^2$, 
while the respective horizon areas are:
$A_{bs} = 4 \pi r_+^2 L$ and $A_{bh} = 2 \pi^2 R_+^3$.
Equating the masses and the horizon areas then yields the transition 
point, 
\eqn\tp{r_+ = {2 \over 3} R_+ = {16 \over 27 \pi} L}
Note that $L \approx 3.53 R_+$, so that the 5-d \BH\ can easily
fit in the compact direction\foot{
As an aside, this relation becomes tighter as we increase the 
extra dimension; for example, carrying out the above calculation
for transition point between 5-d \BH\ $\times {\bf R}^1$ and 6-d \BH\
yields $r_+ = {3 \over 4} R_+ = {3^5 \over 2^{10}} L$, so that
$L \approx 3.16 R_+$, {\it i.e.},\ ${2 R_+ \over L} \approx 0.63 < 1$.  
For high dimensions, this ratio eventually gets ``dangerously'' big:
for $d=10$, ${2 R_+ \over L} \approx 0.91$, and for 
$d=26$, ${2 R_+ \over L} \approx 1.48 > 1$, so the critical \BH\ 
would no longer fit into the extra direction.
However, the actual instability, as calculated numerically by 
\GL, occurs for smaller \BH s than the critical ones, and these do 
seem to fit.}.

Let us now compare this value with the actual onset of instability
as determined by the linearized analysis of \grla.
If we fix $L$ and vary $r_+$, we find that the black string becomes
entropically unfavorable for $r_+ < 0.189 L$,
but it becomes actually unstable only for $r_+  < 0.14 L$.
This shows the important point that while the global entropy
argument can reveal where an instability is allowed to occur,
it does not tell us where it actually does occur.  In other words,
the existence of a more entropically favorable configuration
does not require an instability of the initial configuration.

An alternate formulation of the problem which is more 
analogous to the one we will employ later, is to consider both
the parameters $r_+$ and $L$ fixed, and look for unstable modes 
as perturbations of this spacetime.  Then the criterion for finding
unstable modes is a bit different, since these have to be consistent
with the spacetime, {\it i.e.},\ the wavelength has to be quantized:
$\lambda = {L \over n}$ for some integer $n$.  
Suppose now that we are in the stable regime with $r_+ > 0.14 L$.
Despite the existence of a negative Euclidean mode 
({\it i.e.},\ thermodynamic instability), there is no Lorentzian mode
which would be unstable.  Why?  Simply because we would need
the wavelength of such a mode to be sufficiently large, 
$\lambda > 7.1 r_+$, but this is incompatible with 
$\lambda = {L \over n} < {7.1 \over n} r_+$.

This reiterates the point mentioned earlier, which is obvious for
ordinary Schwarzschild \BH s, that thermodynamic instability
does {\it not} generally imply a dynamical instability.
At first glance, there is no reason to expect them to be related at
all, since the former has quantum roots while the latter concerns
a purely classical evolution; it may even seem rather surprising
that there in fact is a deep relation for the noncompact cases.

\subsec{Black string in a box}

The above example has demonstrated that if the dimensionally reduced
\BH\ has negative specific heat, the corresponding black string may
or may not be unstable, depending on the size of the compact direction.
Let us now consider the converse situation, namely what happens
when the dimensionally reduced \BH\ has positive specific heat.
One way to achieve this is by considering charged \BH s as in \gumi;
but a simpler way is to put the \BH\ in a finite cavity, as done
in e.g.\ \grro.

For noncompact translationally invariant solutions, \grro\ find
that an uncharged black brane in a spherical cavity is classically
unstable if and only if it is locally thermodynamically unstable, 
which provides further supporting evidence of the \GM\ proposal.
The important point to note here is that the presence of a boundary
affects the classical instability.  While this may seem rather odd
if one thinks of the \GL\ instability as pertaining to the horizon,
the reason for this behaviour is that the boundary conditions 
restrict the allowed initial perturbations, just as the geometry 
of the internal space in the previous subsection.

Considering black holes or black branes in a box has many similar 
features to considering these black objects in appropriate 
asymptotically Anti de Sitter spacetimes---after all, the AdS
spacetime has a confining potential which acts like a box of size
given by the AdS radius.
Therefore, we may expect many similar features for these cases.
In particular, as discussed below, the specific heat changes sign
as the ratio of the black hole radius to the AdS radius varies.
We will see in Section 4 to what extend these expectations are
realized.

\newsec{Conjecture for Product Spacetimes}

What we have just seen in the survey of various examples gives us a little 
flavour of what to expect in general. First of all, it is clear that 
a simple criterion based on global entropy arguments 
does not suffice to determine
a \GL\ like instability, as manifested clearly by the  black string example. 
At best, it indicates the presence of a more entropically favourable configuration, 
which may be reachable only by quantum tunneling.
To wit, there could be instanton solutions, but the 
classical fluctuations need not be tachyonic. 

On the other hand, a completely local thermodynamic criterion,
such as that proposed by 
Gubser and Mitra, is also unlikely to hold in  more general settings than those 
considered by the authors, 
namely when the transverse space is not ${\bf R}^n$.
The simple reason is that such a criterion applies only to the 
dimensionally reduced \BH, whereas the \GL\ instability explicitly involves the 
transverse space over which the \BH\ is initially smeared.
Hence, if we wish to consider arbitrary transverse spaces,
the proposed criterion should include the information about them somehow.
In fact, from our preceding discussion, we can guess such a criterion
rather easily.

In particular, 
any viable proposal seems necessarily to involve some features that are 
local (in the thermodynamic sense),
but at the same time, also needs to have some global information, 
in the sense of knowing about the full spacetime.
Taking the Gubser-Mitra criterion as the right local criterion, 
one is then led to propose the following:

Consider a static solution to Einstein's equations (with the
 stress tensor satisfying all the necessary requirements), which is of the 
form ${\bf B }_m \times X^n$, where ${\bf B}_m$ is an $m$-dimensional 
black hole spacetime and $X^n$ is an $n-$dimensional ``transverse'' space; 
namely, 
the horizon has a direct product structure with spatial slice 
${\bf S}^{m-2} \times X^n$. 
If ${\bf B}_m$ has a Euclidean 
Lichnerowicz operator with a negative eigenmode, {\it i.e.} 
$$ \Delta_E({\bf B} ) h_{\mu \nu} = \lambda_E \, h_{\mu \nu}$$ 
with $\lambda_E < 0$, and if the usual Laplacian on $X^n$ has 
eigenfunctions $\phi$ with eigenvalues $- \mu^2$ (which can be discrete), 
{\it i.e.},
$$ \nabla^2(X) \phi = - \mu^2 \, \phi, $$

\noindent
then we expect the onset of instability to occur exactly
when we can saturate the 
Euclidean eigenmode with the `momentum' in the $X^n$ directions. 
Namely, 
{\it the threshold unstable mode will occur when $\mu^2 = - \lambda_E$}. 
Furthermore, we expect that
for lower momenta, $0< \mu^2 < - \lambda_E$, there will be dynamical 
instability manifested by modes growing exponentially in time.

Stated differently, for 
\BH\ spacetimes having a direct product structure, 
not only is it required (in the language of Gubser and Mitra) that there
be an instability of the  thermal ensemble as manifested by 
positive eigenvalues of the Hessian matrix,  but also that the  
restrictions the boundary conditions at `infinity' place on the set
of allowed eigenfunctions in the transverse directions (those given by $X^q$), 
be satisfied. 

Several comments about our conjecture are in order.
First, although
the criterion we propose is a very simple modification of the \GM\ conjecture \gumi,
it applies to a much larger class of solutions. 
In our notation, \gumi\ required the transverse space
$X^n$ to be ${\bf R}^n$, whereas for us it can be anything (subject to the 
field equations).
Second, our proposal holds for all the hitherto considered examples, including the 
compact black string as well as the black string in a box.  Clearly,
the \GM\ conjecture is a special case (since one can always 
choose the mode with arbitrary $\mu^2$ on ${\bf R}^n$, and therefore compensate 
for any $\lambda_E$), so that all these cases are satisfied as well.
From an aesthetic point of view,
 it is nice to note that a sufficient admixture of local and 
global conditions seem to be necessary to cover a larger set of examples.
Our proposal is formulated in a fully geometrical fashion and
takes into account the full spacetime, while at the same time it does not rely
on the global entropy arguments.

As mentioned above, our criterion would greatly facilitate the stability
analysis of \BH\ product spacetimes, since one would no longer need to 
perform the explicit linear analysis.\foot{
Although establishing local thermodynamic instability would be easier still
compared to computing the Euclidean negative mode, we refrain from formulating
our conjecture in that language, because of the prevailing difficulty
in associating the Euclidean negative mode to the local thermodynamic
instability in these more general cases.}
Also, note that apart from dictating where one should expect an instability,
an immediate consequence of our proposal is that any black object,
which can be written in a product fashion as above, with positive
eigenvalues of the Euclidean Lichnerowicz operator,
has to be dynamically stable.

The proposal is most justifiable for the threshold unstable mode.
More interesting, but also more uncertain, aspect relates to 
the dynamical instability.  
Clearly, lacking the simple relation between the Euclidean and the Lorentzian 
picture for dynamical settings, we don't have any clear proof of the 
presence of a growing unstable mode
for any $0 < \mu^2 < - \lambda_E$
(although  it seems to hold in the known cases). 
Finally, while a step forward, our conjecture
 does not address the most general spacetime with horizon.
In particular, there are many interesting examples which cannot 
be written in terms of a direct product.
At present, we do not have any proposal pertaining to these cases,
though we now briefly mention one special class of generalizations:

We have in the above restricted our attention to black holes smeared over 
an internal space homogenously. One could also consider examples wherein 
the black holes are warped over an extra dimension\foot{We would like to 
thank G. Kang for bringing this to  our attention.}. In these examples, too,
the static mode can be decomposed as a part coming from the warping direction 
and a part in the black hole spacetime, {\it i.e.}, we simply choose a 
factorised ansatz for our perturbations. Then the fluctuation equations 
split up, as in the above case, to an eigenvalue problem. However, now the 
analog of the `momentum' from the internal directions which plays the 
role of the Euclidean eigenvalue is determined not in terms of simple 
eigenvalues of the Laplacian on the internal space, but 
in terms of the energy levels of a quantum mechanics problem, where the 
potential is a function of the warp fator. Some examples 
along these lines have been discussed in \kanga, \kangb. Generalizations to 
the cases where we have both warping and homogeneous smearing over an 
internal space should be obvious superpositions of the individual cases.

\newsec{Schwarzschild-$AdS_5 \times \S^5$}

Let us now apply our conjecture to a spacetime which has been outside
the scope of the \GM\ conjecture,
namely the Schwarzschild black hole in $AdS_5$ times a 
five sphere, which is a solution to the Type IIB equations of motion. 
In particular, unlike the previous cases with translational symmetry
discussed in Section 2, this is 
  an example where the solution is neither translationally 
invariant nor is the symmetry group along the transverse directions noncompact.
Rather, since the extra directions are along the 5-sphere, 
the original solution has SO(6) symmetry.  
The metric is 
\eqn\adssch{
ds^2 = -V(r) \, dt^2 + {dr^2 \over V(r)} + r^2 d\Omega_3^2 + L^2 d\Omega_5^2
}
where, for Schwarzschild-AdS$_5$, $V(r) = 1 + {r^2 \over L^2} - {r_+^2 \over r^2}
\left(1 + {r_+^2 \over L^2} \right)$; $r_+$ is the horizon radius and
 $L$ is the radius of both AdS and the 5-sphere.

We will now motivate  our
 present guess of where we would expect the instability to  appear.
The first observation is that the Schw-AdS$_5$ \BH\ is thermodynamically
unstable for $r_+ < {L \over \sqrt{2}}$.  This can be seen either
by considering the Hessian matrix\foot{
In particular, 
$M \sim S^{2/3} \, \( S^{2/3} + L^2 \( {\pi^2 \over 2} \)^{2/3} \)$, 
so that the solution is thermodynamically unstable if
${\p^2 M \over \p S^2} < 0$, {\it i.e.},\ 
$S^{2/3} < {1 \over 2} \(  {\pi^2 \over 2} \)^{2/3} L^2$.
Since $S = {2 \pi^2 \over 4} r_+^3$, this translates into
$r_+ < {L \over \sqrt{2}}$.}, 
or more simply by evaluating the specific heat:
The Hawking temperature is 
\eqn\bhtemp{
T = {2 r_+^2 + L^2 \over 2 \pi r_+ L^2}}
so $C_v \sim {dT \over dr_+} = {1 \over  2 \pi  L^2 r_+^2} (2r_+^2 -L^2)
<0$ if $r_+ < {L \over \sqrt{2}}$,
independent of the details of the compact part of the geometry. 
This in fact matches precisely with the Schwarzschild-AdS \BH\ 
Euclidean mode becoming negative, as conjectured by \hapa\ and
demonstrated explicitly by Prestidge \prestidge.

However, this does not provide the true criterion for the dynamical
instability, as explained in the previous section.
All that the local thermodynamic stability
 calculation shows is that a black hole with
$r_+ > {L \over \sqrt{2}}$ should be dynamically stable.
The point is that the unstable mode simultaneously has to be 
an eigenmode of the Laplacian on the 5-sphere.
In particular, 
we expect the lowest allowed $\l$ (denoting the harmonic on the $\S^5$) 
for which unstable mode can exist
to determine the parameters for the onset of the dynamical instability.

\subsec{Threshold unstable mode}

To wit, in order to establish the existence of a static mode that signals the 
onset of the instability, we shall follow the treatment of \krn, \dewolfe\
to obtain the fluctuation equation about the background solution.
Our starting point is the background corresponding to the smeared 
Schwarzschild AdS black hole 
with the metric given by \adssch. 
In addition, one also has a non-trivial 5-form background. 
We shall 
consider only the fluctuations of the metric on the AdS part of the spacetime.
The fluctuation of the metric components along the five sphere directions are 
scalars from the $AdS_5$ point of view, so these should have
no exponentially growing fluctuations  \qnm.
Our ansatz also assumes that at leading order, as in the black string example 
of Gregory and Laflamme \grla,
 the fluctuations of the RR 4-form potential can be consistently set to 
zero. 

The fluctuation equations can be obtained by expanding the equations of 
motion of Type IIB supergravity about this background.
Since we are looking for metric fluctuations of the AdS part, 
we can restrict ourselves to looking at Einstein's 
equations with indices along the AdS directions.
Adopting the notation of \dewolfe,
we write the full linearized Ricci tensor as 
\eqn\linricci{\eqalign{ 
R_{MN}^{(1)}  = &-{1 \over 2} [ (\nabla^2_x + \nabla^2_y ) h_{MN} +
 \nabla_M \nabla_N 
h_{\ P}^P - \nabla_M \nabla^P h_{PN} - \nabla_{N} \nabla^P h_{PM} \cr
 & - 2 R_{MPQN} h^{PQ} - R_{\ M}^P h_{NP} - R^P_{\ N} h_{MP} ],
}}
where $\nabla^2_x \equiv g^{\mu \nu} \nabla_{\mu} \nabla_{\nu}$ and 
$\nabla^2_y \equiv g^{\alpha \beta} \nabla_{\alpha} \nabla_{\beta}$
are the d'Alambertian operators on AdS$_5$ and $S^5$, respectively.

\noindent
Since the equation of motion is 
\eqn\tendeom{
R_{MN}^{(1)} = {1 \over 48} F_{M P_2 P_3 P_4 P_5} F_{N}^{P_2 P_3 P_4 P_5} 
- { 1 \over 480} g_{MN} F_{(5)}^2,
}
\noindent
the fluctuation equation for the AdS part simplifies to
\eqn\adsfluc{\eqalign{
-{1\over 2} [ & (\nabla^2_x + \nabla^2_y ) h_{\mu \nu} + \nabla_\mu \nabla_\nu
h_{\ \rho}^\rho  - \nabla_\mu \nabla^\rho h_{\rho \nu} - \nabla_\nu \nabla^\rho
h_{\rho \mu} - 2 R_{\mu \rho \sigma \nu } h^{\rho \sigma} \cr
& - R_{\ \mu}^\rho h_{\rho \nu} - R_{\ \nu}^\rho h_{\rho \mu} ] 
+ {4 \over L^2} h_{\mu \nu}  = 0.
}}

\noindent
We make the following choice as our ansatz for metric with the fluctuations:
\eqn\metfluc{\eqalign{
ds^2 = - V(r) & \left( 1 + \epsilon \psi(r) Y_{\l}(\Omega_5) \right) \, dt^2 +
 {1 \over V(r) } 
\left( 1 + \epsilon \chi(r)  Y_{\l}(\Omega_5)\right) \, dr^2 \cr
& + r^2 \left( 1 + \epsilon \kappa(r) Y_{\l}(\Omega_5)\right) 
d\Omega_3^2 + L^2 \, d\Omega_5^2;
}}
without loss of generality, we can set $L \equiv 1$ and subsequently
measure all lengths in terms of the AdS units.
Here $Y_{\l}(\Omega_5)$ denote the usual spherical harmonics on the five sphere, 
satisfying $\nabla^2_y \, Y_{\l}(\Omega_5) = -\l(\l+4) \, Y_{\l}(\Omega_5)$.
In addition we choose a transverse traceless gauge for the fluctuations, 
$$ \nabla^{\mu} h_{\mu \nu} = 0 , \;\;\;\;\; h^{\mu}_{\ \mu} =0, $$

\noindent
which implies
\eqn\gaugecond{\eqalign{
& \kappa(r) = -{1 \over 3 }\( \psi(r) + \chi(r) \) \cr
& \psi(r) = {2 r V \over r V' - 2 V} \, \chi'(r) + { r V' + 8 V \over r V' - 2V}
\, \chi(r).}}

The equation of motion for the $(rr)$ component is:
\eqn\chieq{\eqalign{
-V \chi''(r) + & \left\{ {2 r^2(VV'' -V'^2) - 3rVV' + 10V^2 \over r(rV'-2V)}
\right\} \chi'(r)  \cr
& + \left\{ {r^2 V' V'' + r(8 VV'' -7V'^2) + 4 VV' \over r(rV'-2V)} \right\}
\chi(r) = - \l(\l+4) \, \chi(r)
}}
 
\noindent
where we have eliminated the functions $\psi(r)$ and 
$\kappa(r)$ using the gauge conditions \gaugecond.

\noindent
We in principle have another equation from the $(tt)$ component,
which we present here for completeness: 
\eqn\psieq{\eqalign{
V \psi''(r) + \( V'(r) + {3 V \over r } \) \psi'(r) - 
\( {V'^2 \over 2 V} - {V' \over r} \) \psi(r) - &
\( V'' - {V'^2 \over 2V} - {V' \over r} \) \chi(r) = \cr 
& = - \l(\l+4) \, \psi(r)
}}

\noindent
When we use the second gauge condition, this becomes a third order differential
equation for $\chi(r)$;
however, the solutions of this equation 
are different from those of \chieq\ only by a pure gauge term. 
Similarly, all the components of \adsfluc\ are either trivial or physically 
equivalent to \chieq. 

Hence, to find the metric fluctuations, 
 we can solve eqn.\chieq,
subject to the physical boundary conditions:
regularity everywhere (including the horizon) and 
normalizability given by appropriate fall-off at infinity. 
Having solved for $\chi(r)$, we can then use \gaugecond\ to obtain 
the other components of the metric fluctuation.
This defines an eigenvalue problem: 
for every $r_+$, we can find the value of the 
real (but not necessarily integral) number $\l$, for which there exists a
solution $\chi(r)$ satisfying the boundary conditions.
Alternately, for any fixed $\l$, we can find the corresponding value of $r_+$,
which is physically more relevant for the problem at hand.
Thus, rather than finding the explicit form of the metric fluctuation,
we wish to find the `eigenvalue' $r_+$, for $\l = 1,2,3,...$.

However, before presenting the results, let us first make a few comments about
the $\l =0$ case.
Using the ansatz $L=1,\  \l=0,\  r_+=1/\sqrt{2}$, we can actually find an
analytic solution to \chieq.
The metric fluctuations (as given by \metfluc) are:
\eqn\explsol{
\psi(r) = {2r^2 + 1 \over 2r^4 (2r^2+3)}, \ \ \ \ \ 
\chi(r) = {1 \over r^4 (2r^2+3)}, \ \ \ \ \ 
k(r) = - {1 \over 6 r^4}. }
One can easily verify that this satisfies all the regularity constraints 
and falls off sufficiently fast at infinity.
We may then ask, why is this not the first threshold unstable mode?
The answer is that $\l = 0$ corresponds to the zero mode on the 5-sphere,
so such deformation would not represent a \GL\ instability.
Also, we know that in the Kaluza-Klein reduced picture, AdS \BH s are stable.
One may still worry that, in this 5-dimensional picture, our (linearized)
solution would violate the \BH\ uniqueness; however, one can easily check
that with suitable change of coordinates, it corresponds to the standard
Schwarzschild-AdS$_5$ \BH\ with slightly modified mass.
Hence, we learn that this was a rather special case: Only at this
point did the first variation in temperature with respect to $r_+$ vanish, 
while for all other values of $r_+$ (or equivalently $\l$) the mass remains
fixed even at linear order.

We now present the results for the higher values of $\l$, which 
correspond to genuine threshold unstable modes.
Table 1 lists the values of $r_+$ for $\l = 1, ... , 5$, and we
include  the corresponding Euclidean negative mode, $\lambda_E = -\l(\l+4)$.
%
%
\midinsert
\centerline{%
\vbox{
  \offinterlineskip \tabskip=0pt
  \halign{\strut
          \vrule#&              %
          \hfil $ #~$ &\vrule#& %
          \hfil $\,#$ &         %
        ~ \hfil $#$ &\vrule#&   %
          \hfil $\,#$ &         %
        ~ \hfil $#$ &\vrule#&   %
          \hfil $\,#$ &         %
        ~ \hfil $#$ &\vrule#&   %
          \hfil $\,#$ &         %
        ~ \hfil $#$ &\vrule#    %
          \cr
     \noalign{\hrule}
     \noalign{\hrule}
 & \omit ~$\ell$
       &&  \omit \hfil $\ \lambda_E$ \hfil & \omit \hfil $r_+$ \hfil & \cr
     \noalign{\hrule}
    & \ 1   && -5   &  0.4259   &   \cr
    & \ 2   && -12  &  0.3214   &   \cr
    & \ 3   && -21  &  0.2478   &   \cr
    & \ 4   && -32  &  0.2056   &   \cr
    & \ 5   && -45  &  0.1759   &   \cr
     \noalign{\hrule}
                                                             }}}
\smallskip
{\bf Table 1:}  Values of the black hole radius $r_+$ corresponding
to the threshold unstable mode for the first few harmonics $\ell$.
\endinsert

\noindent
We also plot ${\sqrt{2} \over L} r_+$ as a function of $\l$ 
in Figure 1, where we included a fit to guide the eye better.  
This rescaling makes it easy to see  that as $\l \to 0$,
$r_+ \to {L \over \sqrt{2}}$, which
corresponds to the onset of the local thermodynamic instability.

\ifig\rlplot{For Schw-AdS$_5 \times \S^5$ black holes, 
${\sqrt{2} \over L} r_+$ as a function of $\l$,  
with the corresponding fit to guide the eye.}
{\epsfxsize=9.5cm \epsfysize=5.5cm \epsfbox{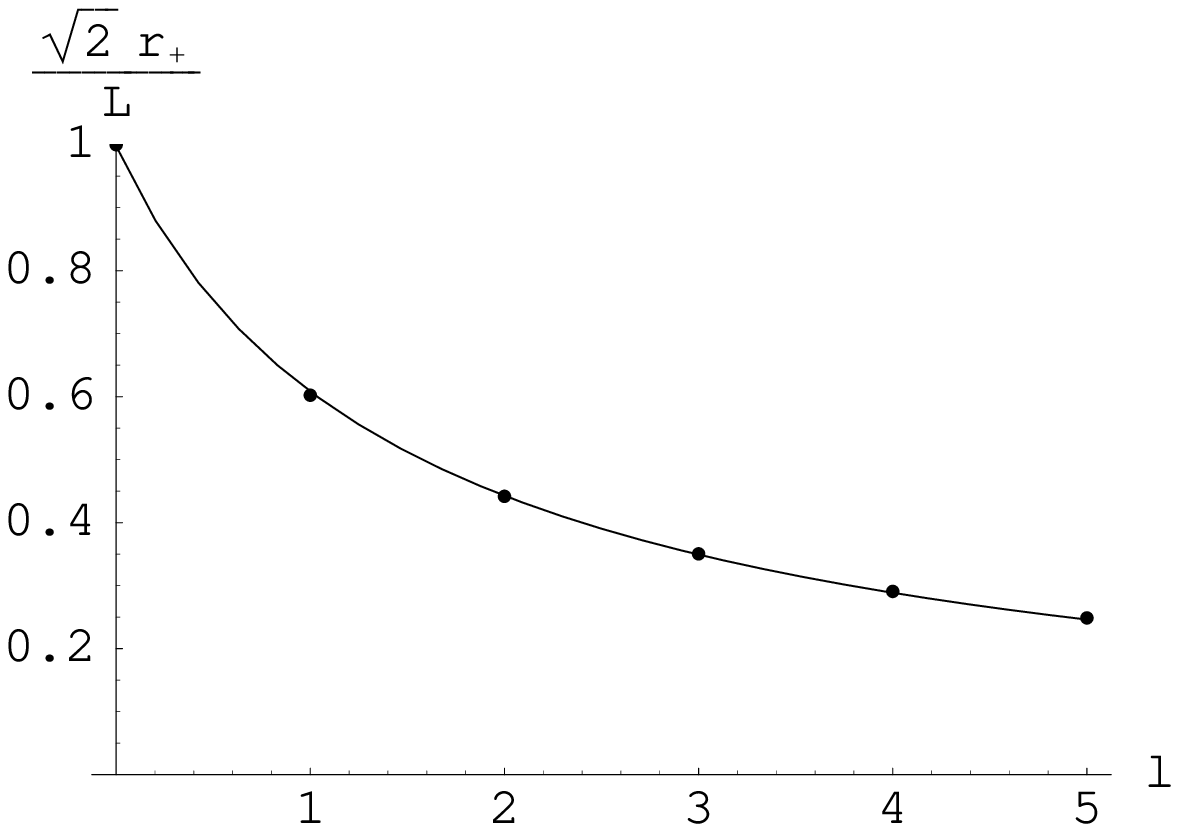}}

Not surprisingly, \chieq\ is
the same equation as that obtained by Prestidge \prestidge, who 
examined the eigenvalues of the Euclidean Lichnerowicz operator. In his 
study, the rhs of \chieq\ had $-\l(\l+4)$ replaced by $\lambda_E$, the Euclidean 
eigenvalue. Prestidge found that for $r_+ > L/{\sqrt 2}$ the eigenvalue 
was positive, at $r_+ = L/{\sqrt 2}$ there was a mode with zero 
Euclidean eigenvalue, and for smaller black holes there was a negative 
eigenvalue. The eigenvalue $\lambda_E$ appeared to be a monotonically 
increasing function of $\rho_+ \equiv { \sqrt{2} r_+ \over L}$, going to 
negative infinity as $\rho_+ \to 0$. 
Furthermore,
\prestidge\ also plotted the Euclidean eigenvalue $\lambda_E$ as a function
of $\rho_+$.
All this agrees with our results, as one can see from Fig.1.

Hence, we have established the existence of the static mode in the Lorentzian 
geometry which marks the separation between the stable and the unstable regime.
We have  verified numerically that the above equation, defining a
well-posed eigenvalue problem, has a solution for various values of $\l$.

\subsec{Gauge theory implications}

Part of the motivation for considering the 
Schwarzschild-$AdS_5 \times \S^5$ solution
of course rests on the AdS/CFT correspondence \refs{\juan,\witten,\gkp,\magoo}.
Let us now consider what all this implies for the 
corresponding dual 4-dimensional SYM gauge theory.

Large \BH s in AdS are described by an approximately thermal state,
with  temperature T given by \bhtemp,
in the gauge theory.
However, small \BH s (compared to the AdS radius)
do not have any role in the thermal ensemble. The reason 
is the following: When we are considering thermal Yang-Mills theory
on $\S^3$, we are instructed to look for five dimensional 
spacetimes whose boundaries are $\S^3 \times \S^1$. There are 
two spacetimes which fit this criterion \hapa, \witheads;
one is the 
Schwarzschild black hole in AdS and the other is a hot thermal gas in AdS.
Which of these geometries contributes dominantly to the field theory 
is decided in terms of their contribution to the free energy as per the 
Euclidean path integral. The black hole contributes dominantly for 
high temperatures, while the low temperature behaviour is governed by the 
thermal gas in AdS space. The transition occurs at $r_+ \sim L$; in the 
field theory side this is best thought of  as a deconfinement transition in 
large N gauge theory \witheads. 
Since the 
thermal gas starts to dominate the saddle point of the 
Euclidean path integral at 
about the point where the black holes in AdS start having negative specific 
heat, it is unclear what the role of the small black holes is in the
thermal ensemble.
This transition is the so-called Hawking-Page transition \hapa, and is
quite distinct from the \GL\ instability which we are interested in.
As in the previous examples, we shall work in  the microcanonical
ensemble, where we can
ask the question about the stability 
of the small black holes in AdS.
 
Let us first consider the small \BH s with ${L/\sqrt{2}} > r_+ > 0.426L$.
While these are locally thermodynamically unstable, as argued above,
they are still dynamically stable.
Although the corresponding state in the gauge theory is no longer the
thermal state, it still maintains the full $SO(6)$ symmetry.
Now we come to the black hole with $r_+ = 0.426 L $.
For such a  black hole we demonstrated that the graviton fluctuation, which is a 
dipole on the $\S^5$, leads to a static threshold unstable mode. This means that 
the dynamical instability wants to deform the homogeneously smeared horizon along the 
$\S^5$ into an inhomogeneous solution, with the horizon shape being 
deformed into one given by the first spherical harmonic. 
Now it is clear what should happen in the dual field theory. 
The SYM$_4$ on $\S^3$ must undergo a phase transition 
to a phase where the $SO(6)$ global symmetry is broken to the subgroup $SO(5)$.
Thus, from the field theory viewpoint, the \GL\ instability is simply 
a symmetry breaking phase transition.

This was argued previously by Gubser and Mitra \gumi;
however, these authors considered only the black brane cases, 
as discussed above.
In fact, the main reason of \gumi\ for considering noncompact symmetries 
(apart from excluding the 
cases which
are  dynamically stable despite being thermodynamically unstable)
rests on the fact that by the AdS/CFT duality,
  the \GL\ instability of a \BH\ horizon
may be viewed as a phase transition in the dual gauge theory; 
but real phase transitions can only occur (or are well-understood)
in noncompact systems, where there are no finite volume effects.
Yet it is by now well-established that the \GL\ instability 
persists even in compact systems, and moreover in gravitational 
systems having a dual field theoretic description. 
Thus, one can ask,  what kind of a transition 
do these instabilities correspond to? 

A related question in this context is, given that there are other 
stable inhomogeneous configurations that are likely candidates for the 
end-point, can one make a statement about what order phase transition 
in the thermal ensemble is one encountering at the onset of the 
instability. One would have thought that the phase transition would 
have to be a second order transition since the existence of the 
threshold unstable mode would imply that one can smoothly deform the 
solution away from the initial configuration\foot{
Note that second order phase transition is not a-priori disallowed
as it would be if the small \BH\ in fact localized on the 5-sphere, 
as was commonly believed until the work of \homa. Since these authors
showed that the horizon can't fragment, but rather stabilizes in some
nonuniform state, this nonuniform state could possibly be smoothly 
matched to the uniform state.}.
However, 
whereas one could make the same argument in the context 
of the \GL\ black string, 
numerical investigations in this 
direction \gubserfo\ 
seem to indicate that for the black string,
the transition is first order.
It would be interesting to understand this issue better, 
particularly for the Schw-$AdS_5 \times \S^5$ case under consideration.

Now what about the higher harmonics $\l > 1$? 
If we consider $r_+ < 0.426 L $, then, according to our conjecture, 
we expect a dynamical instability with the linearized mode (which is still
a dipole on the 5-sphere) growing exponentially in time\foot{
Although we have not done so, it would be an interesting and 
useful exercise to check this numerically.}.
In other words, if we already allow for the $\l=1$ 
mode to be the progenitor of the phase transition, then we no longer have the 
right to condense the other modes. 
The issue of stability of the condensed phase should then be 
discussed from the deformed solution point of view. 
However,  
if we introduce 
some boundary conditions at infinity to disallow the $\l=1$ mode from condensing, 
then we could condense the higher harmonics to break the symmetry. 
This means that the gauge theory should admit an infinite tower
of phase transitions (with appropriately restricted symmetries), 
for smaller and smaller values of $r_+$, the first few of which are given
in Table 1.  
It is perhaps surprising that such a (relatively) simple gauge theory
exhibits such a rich behaviour; but this is just another remarkable
consequence of the AdS/CFT duality.

One could have also considered other internal spacetimes, such as ${\bf RP}^5$ 
or ${\bf T}^{1,1}$, and modulo details depending on the eigenfunctions of the 
internal spacetime, which would affect where exactly the instability starts out, 
we would still have symmetry broken phases for the dual gauge 
theory.

\newsec{Cosmological Horizons}

So far, we have considered only spacetimes with black hole event horizons, 
as defined by the boundary of the causal past of $\cal{I}^+$;
 for  static spacetimes, this is identical to the marginally trapped 
surface, {\it i.e.}, the apparent horizon. Either definition of the horizon
is observer-independent.
But there also are well-known examples of causally nontrivial 
spacetimes with observer-{\it dependent} horizons.  
The simplest and most important of these is the de Sitter (dS) spacetime,
which has a cosmological horizon.
This is a maximally symmetric solution to Einstein's equation with a
positive cosmological constant $\L$.
In  static coordinates, the metric of 4-dimensional de Sitter spacetime
is
\eqn\dS{
ds^2 = - \( 1 - {\L \over 3} r^2 \) \, dt^2 +
       \( 1 - {\L \over 3} r^2 \)^{-1} dr^2 + r^2 \, d \Omega_2^2.}
For any geodesic observer (which in the above coordinates would sit
at $r=0$), the cosmological horizon is centered on the observer, at
radius $r_c = \sqrt{3 \over \L}$.

It is by now well-known that in many respects, cosmological horizons
behave similarly to the black hole horizons; for example, they are
subject to the same laws of black hole mechanics such as the area
theorem.
One might then expect that they could also be dynamically unstable, in a similar
fashion as black hole horizons are.
The main question then is, {\it can observer-dependent horizons be \GL\ unstable?}

Naively, one might expect that the answer should be ``no'', because
of the observer dependence:  
In all the \BH\ examples,
one usually associates the \GL\ instability
with dynamical deformation of the horizon, {\it i.e.},\ the metric
perturbation is largest in a
particular region of spacetime.  Similarly, one usually considers 
the perturbations which are spherically symmetric in the black hole 
directions, around the singular origin.  
In de Sitter, the origin, and consequently the position of the horizon,
is not physically distinguished from any other point; it is rather 
like a gauge choice. 
Thus, one can ask,
how can a physical instability depend
on a ``gauge choice'', namely where the observer sits?
The answer is that it of course can't.
In particular, the candidate metric perturbations 
leading to a \GL\ instability
must preserve the  de Sitter symmetry group.
This can only be done by changing the size of de Sitter.
One in fact can think of this somewhat analogously to considering the spherically
symmetric perturbations of the dimensionally reduced \BH, while the
size of the horizon becomes nonuniform in the transverse directions.
This intuition provides a hint as to what type of perturbations we should
be looking for.

In analogy with our previous discussion of \BH s,
we will focus on spacetimes wherein the cosmological horizon is initially
uniformly smeared over some internal space.
For simplicity, we will work with Einstein gravity with positive cosmological
constant\foot{
This ignores the difficulties string theory has in constructing 
and dealing with such spacetimes, however the examples we'll study here will 
still be of interest from the purely general relativistic point of view.}.
This requires that the ``transverse'' space is positively curved; 
in particular, we can no longer consider $\S^1$ as in the case of the 
\GL\ black string.
A typical candidate would then be, for example, 
$dS_4 \times \S^2$.

For generality, 
we will work in a $d$ dimensional spacetime, dividing $d$ up into a 
de Sitter part of dimension $p$ and an internal part $Y$ of 
dimension $d-p$. The constraint will be that $dS_p \times Y^{d-p}$ 
be a solution to Einstein's equations in $d$ dimensions with fixed
cosmological constant $\Lambda$ {\it i.e.}, solutions to
$R_{MN}-{1 \over 2} R g_{MN} + \L g_{MN} = 0$.  
Taking the trace, we obtain
$R = {2 d \over d-2} \L$,
which implies
\eqn\eeq{
R_{MN} =  {2 \over d-2} \L \, g_{MN}. }
Two interesting $Y$s for us will be 
$\sq$ and $\sqr$, with $q = d-p$ and $q+r = d-p$, respectively.

Let us first consider the metric for 
$dS_p \times \sq$ for general $p$ and $q$.
We can write this as
\eqn\dspsq{
ds^2 = - \( 1 - {r^2 \over L_p^2} \) \, dt^2 +
       \( 1 - {r^2 \over L_p^2} \)^{-1} dr^2 + r^2 \, d \Omega_{p-2}^2
+ \lt^2 \, d {\tilde \Omega}_q^2}
where $L_p$ is the size of the de Sitter part and $\lt$ of the sphere part;
the two being related through 
Einstein's equation 
in terms of $p$ and $q$ as follows:
$L_p^2 = (p-1) \, {(d-2) \over 2 \L}$, and $\lt^2 = (q-1) \, {(d-2) \over 2 \L}$.

Let us now try to see if our idea 
of Euclidean negative modes can be applied here. 
According to our conjecture, if the ``\BH'' (${\bf B}_m$) part of the spacetime
has a negative Euclidean eigenvalue $\lambda_E$, which matches precisely
with the eigenvalue $- \mu^2$ of the Laplacian on the ``transverse'' space
$X^n$, then there exists a threshold unstable mode.
In the present case, we are taking ${\bf B}_m \equiv dS_p$ 
and $X^n \equiv \S^q$.
The Euclidean continuation of 
$p$-dimensional de Sitter is a $p$-sphere; but the Euclidean Lichnerowicz 
operator evaluated in the background of $\sp$ has no negative eigenvalue.
This immediately implies that 
$dS_p \times \sq$ is {\it stable} to the \GL\ clumping on the $\sq$.
Note that we cannot really appeal to any kind of thermodynamic argument 
in this case, since although de Sitter spacetime has a well defined 
entropy, the notion of specific heat is murky since it is hard to 
define the energy of de Sitter space. 

Before proceeding to the second example, let us briefly compare this 
conclusion with what one might naively expect from global entropy considerations.
The initial intuition might lead one to expect the result we found 
above: Einstein's equation fixes the sizes of de Sitter and the sphere
to be comparable, which in the case of the compact black string would
suggest stability.  A more detailed calculation, on the other hand,
shows that $dS_p \times \sq$ is {\it not} the most entropically favorable
configuration if $q \ne 0$.  In particular, the horizon area for
$dS_p \times \sq$ is given by \foot{
This uses the formula for the total horizon area
$A_{p+q} = A_p \times A_q = \omega_{p-2} \, L_p^{p-2} \times \omega_q \, \lt^q$
 where the area of a unit $S^{n-1}$ is 
 $\omega_{n-1} \equiv {2 \pi^{n/2} \over \Gamma(n/2)}$. 
 In the special case where $q=0$, one has to divide the area by 2,
 because if we have no internal space, we want to take the area 
 $\omega_0 = 1$ rather than $\omega_0 = 2$.}
 \eqn\Apq{
 A_{p+q} = 2 \pi^{d/2} \( {d-2 \over 2 \, \L} \)^{d-2 \over 2} \, \,
    {(p-1)^{p \over 2} \over \Gamma\( {p+1 \over 2} \) } \, \,
    {(q-1)^{q \over 2} \over \Gamma\( {q+1 \over 2} \) } .}
One can easily see that the area is maximized for pure de Sitter. 
A global entropy argument would then suggest that $ dS_p \times \sq$
should localize\foot{
Of course, one can apply the arguments of Horowitz and Maeda \homa\ to 
show that, just as for the black string,
 the horizon cannot really fragment.}
into pure $dS_{p+q}$.
As we have seen before, and confirmed above, this argument is 
misleading.

Our failure to 
find the cosmological analog of 
a threshold unstable mode is revealing, in that we can now try to 
engineer a configuration that is more likely to exhibit an instability. 
The important point to notice 
 is that while $\sp$ does not have negative eigenmode, 
$\sp \times \sq$ {\it does}. 
This motivates us to consider $dS_p \times \sq \times \sr$,
with the 
$d (\equiv p+q+r)$-dimensional metric 
\eqn\dspsqsr{
ds^2 = - \( 1 - {r^2 \over L_p^2} \) \, dt^2 +
       \( 1 - {r^2 \over L_p^2} \)^{-1} dr^2 + r^2 \, d \Omega_{p-2}^2
+ \lt^2 \, d {\tilde \Omega}_q^2
+ \lb^2 \, d {\bar \Omega}_r^2}
where the sizes of the three components are  again given by
$L_p^2 = (p-1) \, {(d-2) \over 2 \L}$, $\lt^2 = (q-1) \, {(d-2) \over 2 \L}$, 
and $\lb^2 = (r-1) \, {(d-2) \over 2 \L}$.

We shall now argue that this spacetime 
is unstable. 
In terms of the notation of Section 3, we wish to take
$B_m \equiv dS_p \times \sq$, $X^n \equiv \sr$;
and the trick is to use the fact that the Euclidean space
${\bf S^p} \times {\bf S^q}$ has a negative mode. Physically, this mode 
corresponds to increasing the radius of 
one of the spheres and simultaneously decreasing the radius of the other. 
Let us therefore postulate the following ansatz for 
small fluctuations for the Lorentzian spacetime:
\eqn\pqrfluct{\eqalign{
h_{\mu \nu } & = {1 \over p} g_{\mu \nu} \phi(\bar{\Omega}_r) \cr
h_{a b  } & = -{1 \over q} g_{a b} \phi(\bar{\Omega}_r) \cr
h_{\alpha \beta} & = 0
}}

\noindent
where we use $(\mu,\nu)$, $(a,b)$ and $(\alpha,\beta)$
to denote indices along $dS_p$, ${\bf S^q}$ and ${\bf S^r}$, respectively,
and later $(M,N)$ to denote indices of the full $d$ dimensional spacetime.
The function $\phi(\bar{\Omega}_r)$ 
is a function only of the coordinates on the 
$r$-sphere, because we are only changing the relative radii, but keeping the
de Sitter and $q$-spherical symmetries.

The fluctuations are easily seen to be traceless and 
satisfy the transverse gauge condition $\nabla^M h_{MN} = 0$.
The equation of motion is the linearization of \eeq\ {\it i.e.},
$R_{MN}^{(1)}(h) = {2 \over d-2}\Lambda h_{MN}$ and the linearized Ricci tensor 
is the same as in \linricci. 
With our gauge choice, this simplifies to 
\eqn\linrds{
R_{MN}^{(1)} = -{1 \over 2} 
[(\nabla_\mu \nabla^\mu + \nabla_a \nabla^a + \nabla_\alpha \nabla^\alpha)
 h_{MN} - 2 R_{MPQN} h^{PQ} - R_{\ M}^P h_{NP} - R^P_{\ N} h_{MP} ].}
Using our ansatz \pqrfluct\ the first two terms vanish, and furthermore
it is easy to check that the maximally 
symmetric nature of each of the spaces in the product spacetime causes the 
terms proportional to the Riemann tensor and the Ricci tensor to cancel,
so that the equation of motion reduces simply to

\eqn\pqreqn{
-{1 \over 2} \nabla_\alpha \nabla^\alpha \phi({\bar \Omega}_r) = { 2 \over d-2}
\Lambda \phi(\bar{\Omega}_r).
}

\noindent
One can choose $\phi(\bar{\Omega}_r)$ to be $\l$th spherical harmonic,
with eigenvalue $-{\l(\l+r-1) \over \lb^2}$, and 
using the expression for $\lb^2$, we obtain
\eqn\pqreval{
\l(\l+r-1) = 2 (r-1).
}

\noindent
The only allowed solutions to \pqreval\ are ones where both $\l$ and $r$ 
are integers. A simple solution is $(r,\l) = (2,1)$; in fact, it is also 
easy to check that this is the {\it only} solution for finite $r$.
The establishes for us the threshold unstable mode.

The upshot is that $dS_p \times \S^q \times \S^2$ is {\it unstable} to
small fluctuations, which change the radius of the $dS_p$ and ${\bf S^q}$
commensurately, the functional form of the change being a dipole on the 
${\bf S}^2$. 
Moreover, this is true for all dimensions $p$ and $q$ 
so long as $p+q=d-2$ and $\Lambda$ are fixed.
It is interesting to note that despite the fact that the internal 
space in positively curved, which as \cawa\ argued, would tend
to stabilize the spacetime, we nevertheless find an instability.

Although we have exhibited marginal (static) instability
of  $dS_p \times \S^q \times  \S^2$;
as argued above, we cannot obtain a genuine dynamical instability 
while maintaining the full de Sitter invariance.
However, if we consider perturbations which do break this invariance, 
such argument no longer holds.
Moreover, we can start with cosmological spacetimes which do not respect the
full de Sitter group, such as the Schwarzschild-de Sitter spacetime.
This spacetime now has two horizons, the \BH\ and the cosmological one;
and either one could be subject to the \GL\ instability.
In fact, one might expect that if the \BH\ is dynamically unstable and 
if the \BH\ horizon is sufficiently close to the cosmological horizon, 
then the cosmological horizon must also exhibit dynamical instability,
because otherwise the growth of the \BH\ horizon area without the
accompanying growth of the de Sitter horizon area would threaten to
violate the holographic bound \bousso.
While it is therefore plausible that a cosmological horizon can 
exhibit a dynamical instability, we leave this for future investigation.

\newsec{Discussion} 

One fundamental principle governing the dynamical evolution of 
any system is the second law of thermodynamics, {\it i.e.}, the total 
entropy must be nondecreasing with time.
For isolated black holes, this is identified with the area theorem,
 stating that the 
area of the event horizon cannot decrease in any physically allowed
process.  
While the formulation is quite simple, its consequences are far-reaching.  
One might therefore be inclined to expect that 
an entropy (or horizon area) based argument could
determine the stability of a given solution.

As we saw, there is a hierarchy of entropy-based arguments:
First, the {\it global } argument, such as what \grla\ used to motivate
the instability of a black string, rests on the existence of a
state with higher entropy.
Namely, if there exists a configuration with  entropy higher than that of the
initial configuration, but with all the same conserved quantities, 
then the initial state must be unstable.
However, this criterion is not satisfactory in considering a classical process,
as such a state may be reached from the initial one only via 
quantum tunneling.

This observation led \gumi\ to consider the {\it local } version, 
namely a state would be unstable if the system gains entropy by
locally deforming (usually by ``clumping'') the horizon, while preserving all
the conserved quantities; or
stated more mathematically, if the Hessian of the entropy 
with respect to the extensive variables would have positive eigenvalues.
However, as we argued, even this does not specify the complete picture.
Only for solutions with noncompact translational invariance
 does the conjecture seem to hold!  

In this paper, we have gone one step further, and proposed a criterion
for all classes of static spacetimes which have a direct product structure.
This conjecture has some features of both the local and the global 
arguments.  
While the criterion relates local thermodynamic instability to 
a dynamical \GL\ type instability, as in the \GM\ conjecture,
it is fully geometric and incorporates the information about the 
whole spacetime.
We have checked  a typical and useful example,
the Schwarzschild-AdS$_5 \times \S^5$ \BH,
and found explicitly where the onset of instability occurs.
We have also used our conjecture to motivate an instability of
a cosmological horizon, in the spacetime of the form 
$dS_p \times \S^q \times \S^2$.

However, while we have attempted to address a wider class of solutions, 
we still do not understand the most general types of horizons.
Most importantly, 
we have not addressed the cases where the horizon does
not have a product structure.
While such cases would be interesting and useful to get a handle on,
there are only very 
few solutions which are  known explicitly.
Among these are the $d$-dimensional Schwarzschild and its 
rotating and charged generalizations,
but these are classically stable.
There are also more complicated explicit (stationary)
solutions, such as those considered by
\emre, but they tend to be too complicated to see their stability.
Others, such as the solutions which correspond to the endpoints
that most of the instabilities 
evolve to \homa, or those considered later by \homach,
 are not known explicitly, though we may understand 
some of their properties.

We have been discussing stability of spacetimes 
with horizons.
Part of the motivation  rested on  the fact that 
the special properties exhibited by horizons allow one to 
hope for  a simple  formulation of a universal
criterion for instability.
While we have not been able to address the most general 
static spacetimes with horizons, 
we proposed a generalization of the previous understanding to 
all \BH\ product spacetimes.
We can now look back and ask, to what extent was the presence of a
horizon really necessary for our proposal?
More specifically, one might conjecture that {\it any} product 
spacetime of the form ${\bf B }_m \times X^n$, where 
the ${\bf B}_m$ part of the spacetime
has a negative Euclidean eigenvalue  which matches precisely
with the eigenvalue of the Laplacian on the ``transverse'' space
$X^n$, has  a threshold unstable mode.
While a-priori there seems nothing wrong with this proposal, 
it is difficult to find
(static) spacetimes with negative Euclidean modes 
in the absence of 
horizons.


\vskip .5 cm
\centerline{\bf Acknowledgments}

\vskip .2 cm
It is a pleasure to thank Roberto Emparan, Gungwon Kang, Barak Kol,
Juan Maldacena, Don Marolf, Simon Ross, Matt Strassler, Andy Strominger,
and especially Gary Horowitz for useful discussions.
This work was supported in part by NSF Grants PHY-9870115 and 
PHY-9802484.

%
\listrefs
\end